\documentclass[12pt,fleqn]{article}
\usepackage{psfig,float}
\begin{document}
\begin{center}
{\Large{\bf Study of polarization observables }}
\end{center}
\begin{center}
{\Large{\bf in double pion photoproduction}}
\end{center}
\begin{center}
{\Large{\bf on the proton}}
\end{center}
\vspace{1cm}

\begin{center}
{\large{ J.C. Nacher and E. Oset}}
\end{center}
\vspace{0.4cm}
\begin{center}
{\it Departamento de F\'{\i}sica Te\'orica and IFIC}
\end{center}
\begin{center} 
{\it Centro Mixto Universidad de Valencia-CSIC}
\end{center}
\begin{center}
{\it Institutos de Investigaci\'on de Paterna, Apdo. correos 22085,}
\end{center}
\begin{center}
{\it 46071, Valencia, Spain}
\end{center}

\vspace{3cm}

\begin{abstract}
{\small{Using a model for two pion photoproduction on the proton previously
tested in total cross sections and invariant mass distributions, we evaluate
here polarization observables on which recent experiments are providing new
information. We evaluate cross sections for spin $1/2$ and $3/2$, which are
measured at Mainz and play an important role in tests of the GHD sum rule.
 We also evaluate the proton polarization asymmetry $\Sigma$ which is currently
 under investigation at GRAAL in Grenoble.}}

\end{abstract}
\newpage

\section{Introduction}
Photoproduction of two pions has been the object of intense recent experimental
\cite{bra,za,ha,wolf,metag} and theoretical work \cite{laget,tejedor,tejedor2,
ochi,ochi2,ripani}.
The data are very sensitive to couplings of resonances to photons and mesons and
in particular are a unique source of information on the
$N^\ast(1520)\rightarrow\Delta\pi$ transition which can be contrasted with quark model
predictions \cite{letter}.

The total cross sections and invariant mass distributions bare much information
on the reaction mechanisms and they pose important constraints on the
theoretical models. Yet, further constraints are to be found in the polarization
observables and so far the theoretical models have not tackled this problem. The
advent of recent experiments on this issue makes the theoretical problem
opportune. Furthermore the spin $1/2$ and $3/2$ $\gamma p$ cross sections are
input for the Drell-Gerasimov-Hearm sum rule, which is also receiving much
attention recently \cite{dre,aren}. Given the fact that the photonuclear 
excitation of the proton gives rise
to a rich spectrum of resonances in the $1-2$ GeV region, the $DGH$ sum rule
establishes an interesting link between a static property of the nucleon
and the dynamical mechanisms of the nucleon excitation.

The sum rule was first derived by Gerasimov and by Drell and Hearn in an
independent way and it is based on the work of the Low Energy Theorem of Low
and Gell-Mann and Goldberger for spin $1/2$ particles.

The most important fact of this relation lies in the fact that it is based on 
 general principles as Gauge and Lorentz invariance, cross symmetry,
 causality and unitarity. However, testing the $DGH$ sum rule has proved 
 so far problematic for lack of
data at high energies. This handicap has been overcome by using theoretical
predictions for the one pion photoproduction \cite{dre}. However, since one also
needs the two pion production cross section, the use of theoretical models to
evaluate the contribution of this part becomes also necessary.

At the moment the rough estimates of \cite{karliner} are used in the test of the
$DGH$ sum rule, but the existence of fair models for two pion photoproduction
makes the evaluation of the spin cross sections needed in the $DGH$ test most
advisable. In fact, lack of data or theoretical predictions for the two pion
production on deuterium was the reason in \cite{aren} to stop the integration at
the value of 550 MeV before the two pion production becomes relevant.

The recent measurement of the helicity $1/2$, $3/2$ cross sections in two pion
photoproduction at Mainz \cite{lang} represents an important step in the test
of the $DGH$ sum rules while at the same time it imposes new constraints on the
theoretical models of two pion photoproduction. The present status of the theory
has also experienced a step forward with the solving of the puzzle of the
$\gamma p\rightarrow n\pi^+\pi^0$ cross section which was underpredicted in 
\cite{laget,tejedor2}. The inclusion of the $\Delta(1700)$ excitation, together
with $\rho$ production, lead in \cite{nacher} to good predictions for cross
sections and invariant mass distributions of the $\gamma p \rightarrow
n\pi^+\pi^0$ reaction, without spoiling the agreement found for the other charge
channels. The model of \cite{nacher} seems thus most suited to evaluate the spin
cross sections, hence serving the double purpose of further testing model, while
at the same time, using it as input to test the $GDH$ sum rule.

Additionally more polarization observables, like the polarization asymmetry
$\Sigma$ are been measured in the European detector GRAAL at Grenoble. 
 The perfect cylindrical
symmetry of the GRAAL detector is ideal to measure $\Sigma$. 
Up to now the GRAAL collaboration has presented results for only one emitted
meson in the final state \cite{ajaka} but recently preliminary results for the
$\Sigma$ in the 
$\gamma p\rightarrow\pi^0\pi^0 p$, $\gamma p
\rightarrow\pi^+\pi^-p$ and $\gamma p\rightarrow\pi^+\pi^0 n$
channels are in progress in the range of photon energies 500-1100 MeV
\cite{hourany}.

 With all this information ready to appear in experimental publications in a
short time we will use our two pion photoproduction model to analyze the
observables described above which pose a new challenge to the model.

\section{Helicity asymmetries for  $\gamma p\rightarrow\pi^+\pi^- p$ and
$\gamma p\rightarrow\pi^+\pi^0 n$}

The helicity cross sections $\sigma_{3/2}$ ($\sigma_{1/2}$) are defined as the total cross
section for the absorption of a circularly polarized photon by a proton polarized
with its spin parallel (anti parallel) to the photon spin.

The polarization vectors for circularly polarized photons are :

\begin{equation}
\vec{\epsilon}\, ^{(\pm)} = \frac{(\mp1,-i,0)}{\sqrt{2}}
\end{equation}
By writing our $\gamma p\rightarrow\pi\pi N$ amplitude as $\epsilon_{\mu}\, 
T^\mu$, we evaluate the amplitudes for scattering of the polarized photon with a
proton with spin third component 1/2. Then we have the $T_{1/2}$ and $T_{3/2}$
helicity amplitudes as:

\begin{equation}
T_{3/2}= \frac{-T^x-iT^y}{\sqrt{2}}\, ,
\end{equation}
\begin{equation}
T_{1/2}= \frac{T^x-iT^y}{\sqrt{2}}\, .
\end{equation}

Alternatively, we could also use the photon with $\vec{\epsilon}^{\, (+)}$
polarization and third components $1/2$ or $-1/2$ for the proton spin, as is
customarily done to define helicity amplitudes of resonances.
For a $N^\ast$ resonance the helicity amplitudes $A_{1/2}$ and
$A_{3/2}$ are defined as
\begin{equation}
A_{1/2}^{N^\ast} \sim\langle
N^\ast,J_z=1/2|\vec{\epsilon}\, ^{(+)}\cdot \vec{J}|N,S_z=-1/2\rangle
\end{equation}

\begin{equation}
A_{3/2}^{N^\ast} \sim\langle
N^\ast,J_z=3/2|\vec{\epsilon}\, ^{(+)}\cdot \vec{J}|N,S_z=1/2\rangle
\end{equation}

In these cases $A_{1/2}$ means an incoming nucleon with spin projection
$S_z=-\frac{1}{2}$ (positive helicity) absorbing a photon with spin $\lambda=+1$,
leading to $J_z = \frac{1}{2}$ for the resonance final state (same helicity as in
initial state). For the case of the $A_{3/2}$ it means that we have an initial nucleon spin
projection state
of $S_z=\frac{1}{2}$ (negative helicity) and a photon with $\lambda=+1$, being
the final spin state $J_z=3/2$ (positive helicity, helicity change).


As we said in the Introduction, experiments about that kind of observables and
being performed with the DAPHNE detector at Mainz. The DAPHNE angular acceptance 
for the two charged pion production in the
GDH-Experiment at MAMI for hydrogen target and butanol target is defined as
\cite{lang}
\\

Polar angle$(\theta)$:  $23\leq\theta\leq 158 $\, [deg]
\\

Azimuthal angle$(\phi)$:  $0\leq\phi\leq 360 (2\pi)$\, [deg]  
\\


For the $\gamma p\rightarrow\pi^+\pi^- p$ reaction they can
see up to three charged particles inside the DAPHNE-detector. Depending upon the
number of the particles seen, it is necessary to apply different methods of 
analysis
 which affect the DAPHNE-acceptance for this reaction. The number of
particles has to be handled as follows:

{\it i)}\,  0 and 1 charged particle in the DAPHNE angular acceptance:

The events gets rejected.

{\it ii)}\,  2 charged particle in the DAPHNE angular acceptance

There are two possible sets of particles that can be seen by the 
 DAPHNE detector
in such a case, which are ($p_\pi^{\, \pm}$) or $(\pi^+\pi^-)$.






{\it iii)}\,  3 charged particle in DAPHNE angular acceptance:

In this case the proton momentum threshold ($p_{prot}/(MeV/c)$) against the polar angle
($\theta$ /degrees) is specified 
by the following function:

\begin{equation}
P_{prot}(threshold) > 300 + 0.010(\theta - 90)^2\, [MeV/c]\, ,
\end{equation}

and the charged pion ($\pi^+$, $\pi^-$) momentum threshold ($p_{\pi^{\pm}}$ /(MeV/c))
against the polar angle ($\theta$ /degrees) is specified by:

\begin{equation}
 p_{\pi^{\pm}} > 65 + 0.005(\theta - 90)^2\, [MeV/c]\, .
\end{equation}
 
We adapt our calculations to the $\gamma p\rightarrow\pi^+\pi^- p$ reactions 
are to the experimental acceptance discussed above for the three particle
detection case. In our calculations we use the improved model for two pion
photoproduction mentioned in the introduction \cite{nacher}.
We analyze the $\sigma_{1/2}$, 
$\sigma_{3/2}$ observables and the helicity asymmetry in the following figures.

In fig. 1 we show the $\sigma_{3/2}$ and 
$\sigma_{1/2}$ cross sections for the $\vec{\gamma}\vec{p}\rightarrow p
\pi^+\pi^-$ reaction with the 3 charged particles in the DAPHNE acceptance.
We see that the agreement with the experimental points is quite good 
for
$\sigma_{1/2}$ and $\sigma_{3/2}$. We reproduce a peak around
700 MeV shown by the experimental data. We see a small discrepancy with 
the experiment at photon energies up to 700 MeV 
but the decrease of the cross section up 800 MeV is reproduced.
\newpage
\begin{figure}[h]
\centerline{\protect
\hbox{
\psfig{file=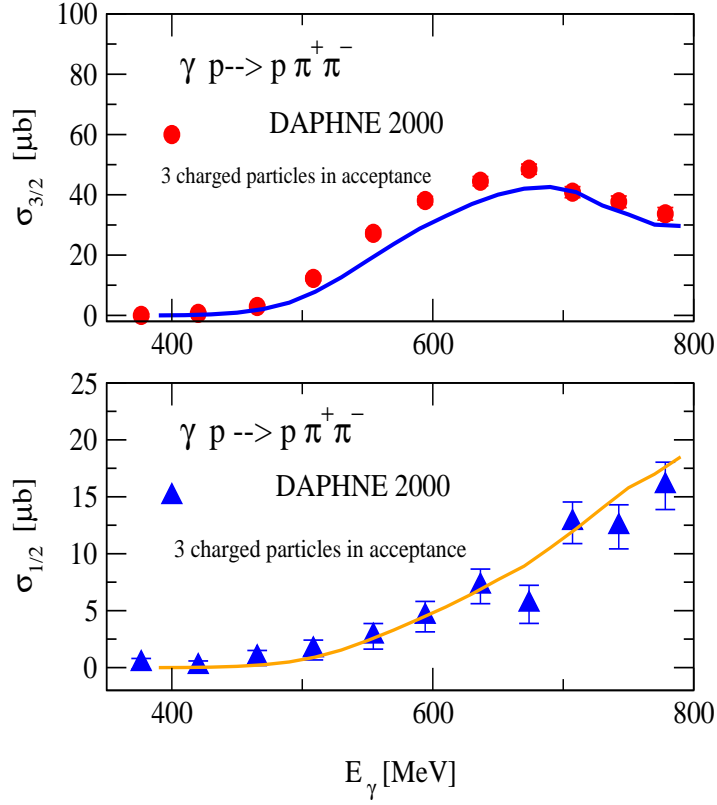,height=15cm,width=17cm,angle=-90}}}
\caption{\small{$\sigma_{3/2}$, $\sigma_{1/2}$ for 
$\vec{\gamma} \vec{p}\rightarrow p\pi^+\pi^-$ with
3 charged particles in DAPHNE acceptance. 
Experimental data from \cite{lang}}}
\end{figure}

In fig. 2 we show the difference between the helicity cross sections
$\sigma_{3/2}$ and $\sigma_{1/2}$ at the top of the figure. In the bottom we
show the helicity asymmetry for the double charged pion reaction with 3
particles in the DAPHNE
acceptance.  We find a similar result for the $\sigma_{3/2}$-$\sigma_{1/2}$
cross section as we showed before for the $\sigma_{3/2}$, reflecting the
dominance of the helicity amplitude $A_{3/2}$. 
The helicity asymmetry is defined by $\frac{\sigma_{3/2} - 
\sigma_{1/2}}{\sigma_{3/2} + 
\sigma_{1/2}}$ and we observe a nice 
agreement with the experimental data although one has large experimental
error bars at low and high energy.

\newpage
\begin{figure}[h]
\centerline{\protect
\hbox{
\psfig{file=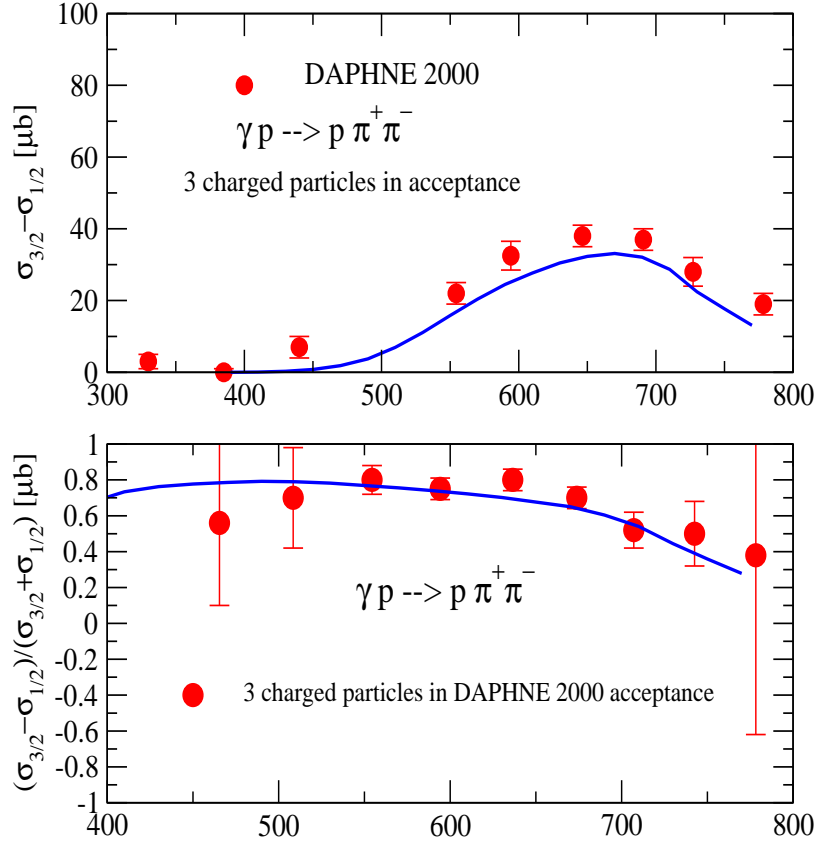,height=15cm,width=17cm,angle=-90}}}
\caption{\small{ The difference of cross sections, 
$\sigma_{3/2}$ - $\sigma_{1/2}$, for the  
$\vec{\gamma}\vec{p}\rightarrow p\pi^+\pi^-$ reaction with
3 charged particles in DAPHNE acceptance, is shown in the top figure  
and the helicity 
asymmetry in the bottom one.  
Experimental data from \cite{lang}}}
\end{figure}

We turn now to the $\vec{\gamma}\vec{p}\rightarrow\pi^+\pi^0 n$ reaction.
In our work \cite{nacher} we added to the model some new ingredients improving 
considerably the results for the $\gamma p\rightarrow n \pi^+\pi^0$ channel 
obtained before in \cite{tejedor2}. We shall show now the results of this
channel for the helicity observables analyzed above.

The DAPHNE acceptance for $n\pi^+\pi^0$ is as follows: the $n$ and 
$\pi^0$, both have no limits in angular acceptance and in threshold momentum.
Only the $\pi^+$ is limited by the angle and momentum threshold. So, in the
$\gamma p\rightarrow n\pi^+\pi^0$ reaction we should take the $\pi^+$ inside the
DAPHNE acceptance. In this case there is a threshold 
for the momentum of the charged pion given as a function of the polar angle
$(\theta/degrees)$ by:
\begin{equation}
 p_{\pi^+} > 80 + 0.005(\theta - 90)^2\, [MeV/c]
\end{equation}

\begin{figure}[h]
\centerline{\protect
\hbox{
\psfig{file=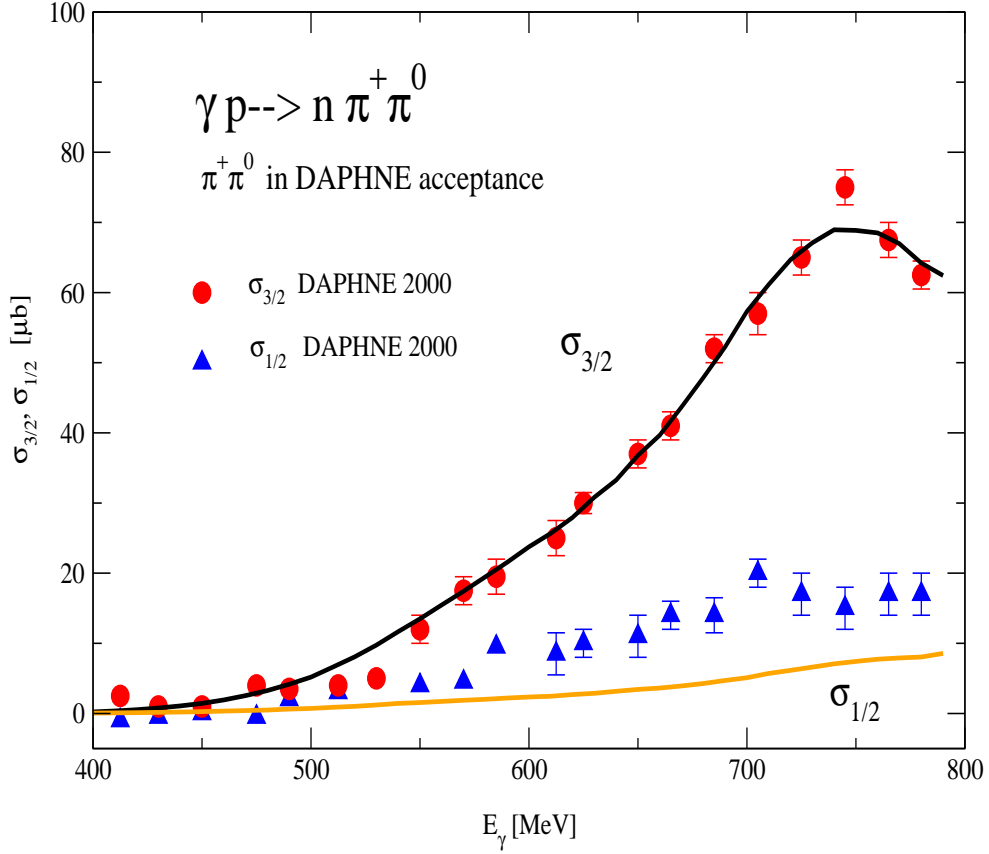,height=14cm,width=15cm,angle=-90}}}
\caption{\small{$\sigma_{3/2}$, $\sigma_{1/2}$ for 
$\vec{\gamma} \vec{p}\rightarrow n\pi^+\pi^0$ with
$\pi^+\pi^0$ in the DAPHNE acceptance. 
Experimental data from \cite{lang}}}
\end{figure}

In fig. 3 we show the results for the helicity cross section 
$\sigma_{3/2}$ with a dark continuous line, which are in good agreement
with the experimental results. In the case of the $\sigma_{1/2}$ we show the
results in a light continuous line and they are somewhat smaller than the
experimental numbers.

In view of these results the small deficit of the theoretical cross section in
this channel found in \cite{nacher} should be attributed to the smaller
theoretical cross section found for $\sigma_{1/2}$.
\newpage
\begin{figure}[h]
\centerline{\protect
\hbox{
\psfig{file=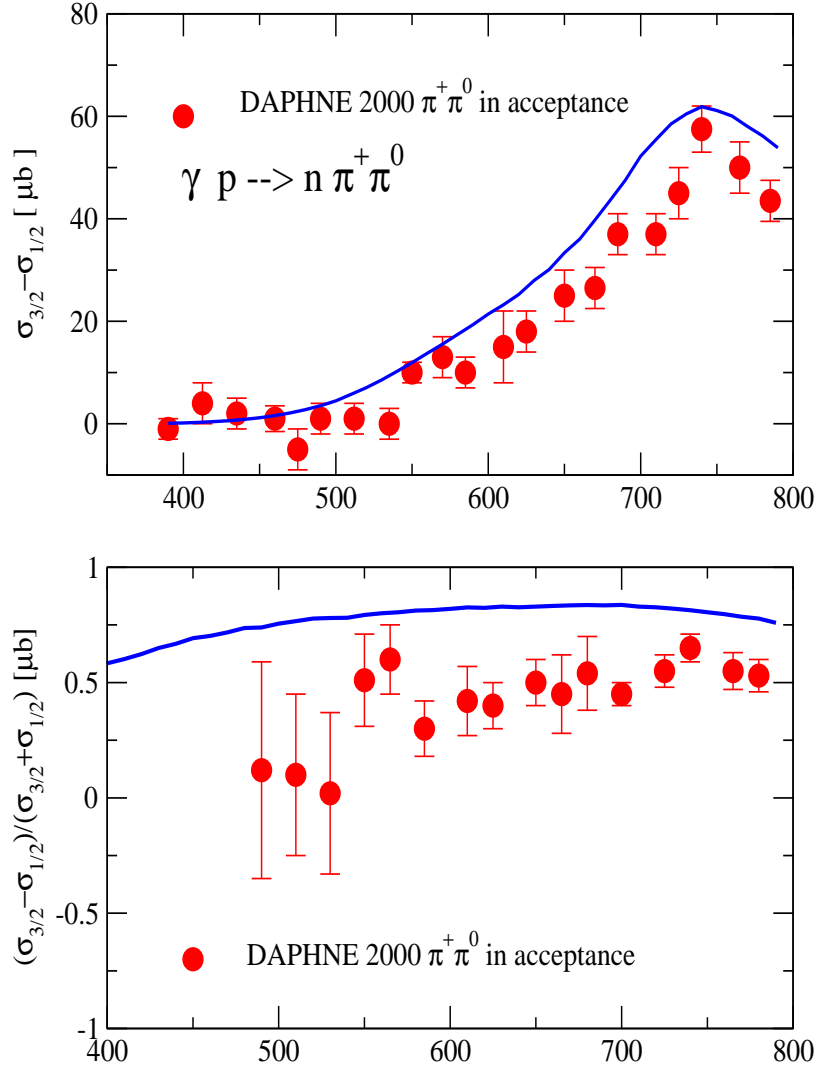,height=15cm,width=17cm,angle=-90}}}
\caption{\small{ The difference of cross sections $\sigma_{3/2}$ -
$\sigma_{1/2}$, for the  
$\vec{\gamma} \vec{p}\rightarrow n\pi^+\pi^0$ reaction with
$\pi^+\pi^0$  in DAPHNE acceptance, is shown in the top figure and the helicity
asymmetry in the bottom one.   
Experimental data from \cite{lang}}}.
\end{figure}

In fig. 4 we show the difference between the helicity cross sections
$\sigma_{3/2}$ and $\sigma_{1/2}$ at the top of the figure. In the bottom we
show the helicity asymmetry for the $\vec{\gamma} \vec{p}
\rightarrow n\pi^+\pi^0$ with $\pi^+\pi^0$
particles in the DAPHNE
acceptance.  We find good agreement for the $\sigma_{3/2}$-$\sigma_{1/2}$
cross section but we realize that if our $\sigma_{1/2}$ helicity amplitude
was a little bit bigger the agreement would be better.
At the bottom  of fig.4 we can see the results for the helicity asymmetry.
We find agreement with the sign of the asymmetry and the global behavior of the
observable 
but the
strength of our results has a discrepancy of about a $10\%$ with the data.

\section{The GDH Sum Rule for $\gamma p\rightarrow\pi\pi N$}

The Drell-Hern-Gerasimov (DGH) sum rule relates
the helicity structure of the photo absorption cross section to the anomalous
magnetic moment. Our aim is to study the convergence of this sum rule in the two pion
charged photoproduction reaction.

We write the DHG sum rule (Drell and Hearn, 1966 \cite{drell}; Gerasimov, 1966
\cite{gera}) as:

\begin{equation}
\frac{\kappa^2}{4}=\frac{m^2}{8\pi^2\alpha}\int\frac{d\nu}{\nu}[\sigma_{3/2}
(\nu)-
\sigma_{1/2}(\nu)] = I^{GHD}(Q^2=0)
\end{equation}
and $\nu=E_{\gamma}^{lab}$ photon lab energy. We studied in the last section the
helicity observables needed to calculate this expression. We also note that the 
vector polarizability may be related to another sum rule as \cite{dre}:

\begin{equation}
\gamma=\frac{1}{4\pi^2}{\int\frac{d{\nu}}{\nu^3}(\sigma_{3/2}-\sigma_{1/2})}
\end{equation}

It is interesting to keep in mind that the helicity cross sections are related
to the total transverse ($\sigma_T$) and transverse-transverse
($\sigma_{TT^\prime}$) cross sections.

\begin{equation}
\sigma_T=\frac{\sigma_{3/2}+\sigma_{1/2}}{2}
\end{equation}
\begin{equation}
\sigma_{TT^\prime}=\frac{\sigma_{3/2}-\sigma_{1/2}}{2}
\end{equation}

We  show in the fig. 5, from up to down, the difference of the helicities 
$\sigma_{3/2}$-$\sigma_{1/2}$
and the integrand of the GHD sum rule in
terms of the laboratory photon energy. 
In the intermediate figure we show the integrand of the $I^{GHD}$ sum rule as a
function of the photon energy. Finally,
we also  show a figure at the bottom of fig. 5 the $I^{GHD}$ in terms 
of the upper limit of photon energy used in the  $I^{GHD}$ integral,
which seem to indicate that the integral does not converge.

We notice that our
model was developed for working in the Mainz range up to 800 MeV 
 photon energy. The result of the $I^{GHD}$ integral up to 800 MeV, 
lower energy where the model is more reliable, is 0.13 for the $\gamma
p\rightarrow\pi^+\pi^-p$ reaction.
We should note that the value quoted in \cite{dre}, for the contribution of the
$\gamma p\rightarrow 2\pi N$ reactions to the $I^{GHD}$ sum rule in 
\cite{dre} from \cite{karliner} is 0.20, counting all the charged channels.
It is worth noting that if we consider the other channels we obtain
a contribution from the $\gamma p\rightarrow\pi^+\pi^0 n$ channel of
0.07 and from $\gamma p\rightarrow\pi^0\pi^0 p$ channel of 0.02 in both
cases integrating up to 800 MeV. The total contribution of the
$(\gamma, 2\pi)$ channels to the $I^{GDH}$ sum rule up to 800 MeV is
0.22, already larger than the previous estimates, which in principle would
account for the integration up to infinite. As an indication of how the GHD sum
rule
might converge we extrapolate our model up to 1 GeV where the agreement with
data should still be fair. We find a contribution to the integral of 0.17, 0.09,
0.02
from the $\pi^+\pi^-$, $\pi^+\pi^0$ and $\pi^0\pi^0$ channel respectively. The
sum of all channels is 0.28, which seems an increase of 20 $\%$ with respect
to the integral up to 800 MeV alone.

\begin{figure}[h]
\centerline{\protect
\hbox{
\psfig{file=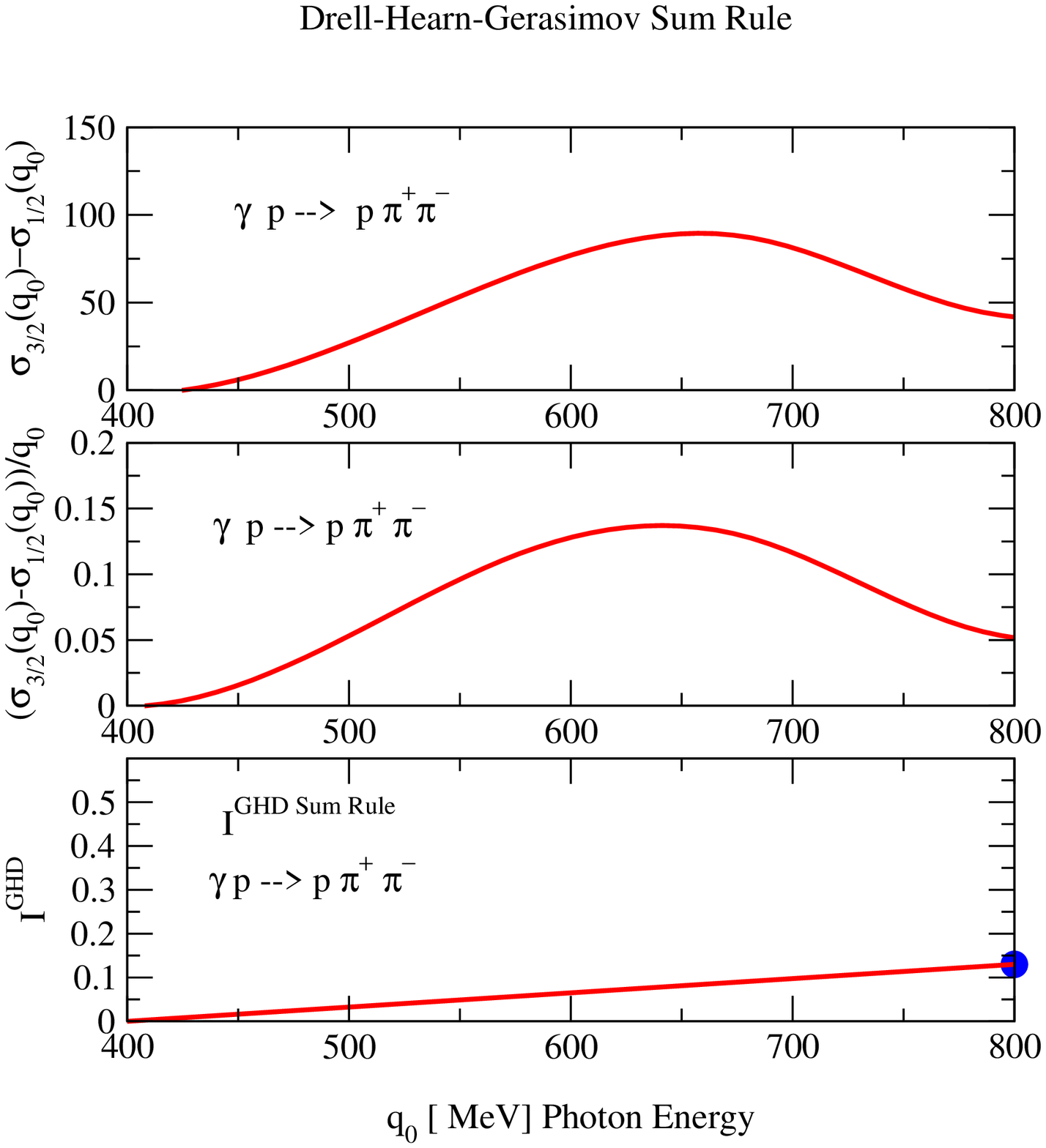,height=12cm,width=15cm,angle=-90}}}
\caption{\small{At the top we show the helicity function $\sigma_{3/2}(q_0)
-\sigma_{1/2}(q_0)$ in terms of the photon lab energy $q_0$ up to 800 MeV for the
$\gamma p\rightarrow\pi^+\pi^- p$ reaction. In the middle figure we show the 
integrand of the $I^{GHD}$ sum rule $[\sigma_{3/2}-\sigma_{1/2}]/q_0$ up to 800
MeV. At the bottom we show the $I^{GHD}$ sum rule in terms of the photon energy.
}}
\end{figure}

We hesitate to use our model at higher energies where it has not been tested
against experiment, and where we know that consideration of further resonances
and extra unitarity corrections should be important. But the results for the 
GHD integral shown in fig. 5 and the increase found from 800 MeV to 1000 MeV, do
not go in the direction of supporting a fast convergence of this integral.
Actually the possibility that the GHD integral does not converge was
already advanced in \cite{dretjnaf}.

As with respect to the sum rule for the vector polarizability the fact that we
have two extra powers of $\nu$ in the denominator increases the chances of
convergence. 
We show in fig. 6 the integrand of eq. (10) for the three charged pion channels
and at the bottom of the panel we show the contribution of the integral in terms
of the upper limit of integration. The convergence of the
$\gamma p\rightarrow \pi^+\pi^-p$ and $\gamma p\rightarrow \pi^0\pi^0 p$
channels is not fast either and the integral does not seem to saturate around 800 MeV. Furthermore, the
$\gamma p\rightarrow \pi^+\pi^0 n$ channel shows even a slower convergence. 
Indeed if we integrate up to 800 MeV the values obtained for the integral are
$8.6\cdot 10^{-5}$ $\mu b/MeV^2$, $4.2\cdot 10^{-5}$ $\mu b/MeV^2$, 
$1.3\cdot 10^{-5}$ $\mu b/MeV^2$ from the $\pi^+\pi^-p$, $\pi^+\pi^0n$,
$\pi^0\pi^0p$ channels
respectively. Altogether we find a contribution to 
$\gamma$ from the two pion production channels of 0.00014 $\mu b/MeV^2$  up to
800 MeV.
If we integrate up to 1000 MeV then the results are 
$1.0\cdot 10^{-4}$ $\mu b/MeV^2$, $6.4\cdot 10^{-5}$ $\mu b/MeV^2$, 
$1.7\cdot 10^{-5}$ $\mu b/MeV^2$
respectively. The sum of all channels up to 1 GeV is 0.00018 $\mu b$/MeV$^2$.
However, in spite of this apparent lack of convergence, and admitting that our results at larger energies
should overestimate the data, we still find a convergence of the integral around
2 GeV with a value of 0.00030 $\mu b$/MeV$^2$, which we believe is an
overestimate of the actual result. Hence we can safely put the results of the $2\pi$ channels to the
$\gamma$ sum rule in the range of [0.00018, 0.00030] $\mu b$/MeV$^2$. 

\begin{figure}[h]
\centerline{\protect
\hbox{
\psfig{file=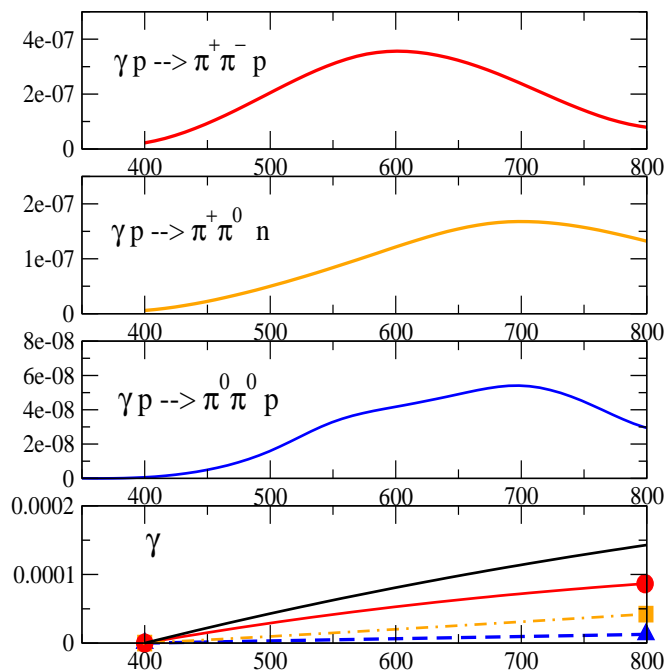,height=12cm,width=14cm,angle=-90}}}
\caption{\small{We show the integrand of the vector polarizability $\gamma$
for the three charged pion channels. At the bottom we show the $\gamma$ sum rule in terms of the photon energy.
for the two pion production channels: (circles) $\pi^+\pi^-$, (squares)
$\pi^+\pi^0$, (triangles) $\pi^0\pi^0$. The upper continuous line shows the total
sum for the vector polarizability sum rule coming from the three channels.}}
\end{figure}

\section{Polarization asymmetry $\Sigma$ for two pion photoproduction}

For photons linearly polarized in the vertical plane with a polarization
degree $P$, the differential cross section can be written as
\begin{equation}
(\frac{d\sigma}{d\Omega})_{pol}(\theta, \phi)=(\frac{d\sigma}{d\Omega})_{unpol}
(\theta)\, (1+P\Sigma(\theta)  cos(2\phi))\, ,
\end{equation}
where $\phi$ is the angle between the reaction plane and the horizontal plane
and $\Sigma(\theta)$ is the beam asymmetry. The cylindrical symmetry of the detector
provides the distributions of selected events over the full range $0-360$
$[deg]$ of $\phi$ angles \cite{ajaka}.

The asymmetry is extracted experimentally from the azimuthal distribution of
events for one of the polarization states, normalized to the azimuthal
distribution corresponding to an unpolarized beam.

\begin{equation}
\hspace{-1cm}
(\frac{d\sigma}{d\Omega})_{pol}(\phi, \theta)/(\frac{d\sigma}{d\Omega})_{unpol}
(\theta)
=1+P\Sigma(\theta)  cos(2\phi)= \frac{2 F_ {ver}(\phi)}{(F_{ver}(\phi) 
+ \alpha F_{hor}(\phi))}
\, , 
\end{equation}
where $F_{hor}$ and $F_{ver}$ are the measured azimuthal distributions of events
for each polarization state and $\alpha$ is the ratio of beam fluxes corresponding
to the vertical and horizontal polarizations. A fit to the experimental data
using the function
$1+P\Sigma(\theta) cos(2\phi)$ allows to extract the beam asymmetry
$\Sigma=\Sigma(E_{\gamma},\theta_{c.m.})$, as a function of the photon energy and
the polar angle $\theta$ in the centre of mass of the selected particle.

In our case, we implement the calculation of the beam asymmetry $\Sigma$
in our model with the following prescriptions:
\begin{itemize}
\item 
The photon momenta is taken in the positive $z$ direction and the $x$ and $z$
axes define the horizontal plane. The vertical polarization goes along the
positive $y$ axis and the horizontal polarization along the $x$ axis.

\item
The reaction plane which contains the momenta of the final particles, 
defines an azimuthal angle $\phi_R$ with respect to the 
initial plane, rotating around the direction $z$ of the incoming photon ($z$
axis).
\item
Since the only $\phi$ dependence of the cross section comes from the
factor $cos(\phi)$ in eq. (13), we evaluate $\Sigma(\theta)$ choosing $\phi=0$.
For this purpose we select small $\phi_R$ angles, thus having the final
particles essentially in the horizontal plane.
\item The asymmetry $\Sigma$ in our case is defined by
\begin{equation}
\Sigma=\frac{\sigma_T - \sigma_L}{\sigma_T + \sigma_L}
\end{equation}
where $\sigma_T$ corresponds to photons polarized along the $y$ axis while
$\sigma_L$ corresponds to photons polarized in the $x$ axis.
\item
In a $\gamma p\rightarrow N \pi_1 \pi_2$ reaction
we consider two kind of plots for each channel. We calculate the beam 
asymmetry for the emission of the system of two pions
($\pi_1 \pi_2$) against the $\theta_{c.m.}$ of the two pions in the global
${C.M.}$ (case $\gamma p\rightarrow\ N\,  (\pi_1\pi_2)$). In this case we will
show the results selecting the peak of invariant mass of $(\pi_1 \pi_2)$ system 
for
several cuts and without cuts.

\item
The other case is where we calculate the beam asymmetry for the $\pi_1$ (or $\pi_2$)
 emission against the $\theta_{c.m.}$ of $\pi_1$ ($\pi_2$) in the global $CM$ of
 the reaction (case $\gamma p\rightarrow\  (N\pi_2)\, \pi_1$) or
 ($\gamma p\rightarrow\  (N\pi_1)\, \pi_2$). In these situations we
 select the peak in the invariant mass of the $(N\pi$) system  around the masss of
 the
 $\Delta$ resonance. The results without these cuts are also analyzed.
This $\theta_{c.m.}$ is defined by the angle between the photon ($z$ direction) and the momentum of
the emitted selected particle.

The selection of cuts discussed above are set up for a future comparison with
data presently been taken at GRAAL \cite{hourany}.
\end{itemize}

In the next pages we can see our predictions for these observables.
From fig. 7-13 we analyze the three possible isospin channels with a proton in
the initial state, $\gamma p\rightarrow\pi^+\pi^-$, $\gamma p\rightarrow\pi^0\pi^0 p$,
and $\gamma p\rightarrow\pi^+\pi^0 n$ reactions. We show the beam asymmetry in
several cases, selecting one or two pions and with or without cuts in the
invariant masses of the systems.
\newpage

\begin{figure}[h]
\centerline{\protect
\hbox{
\psfig{file=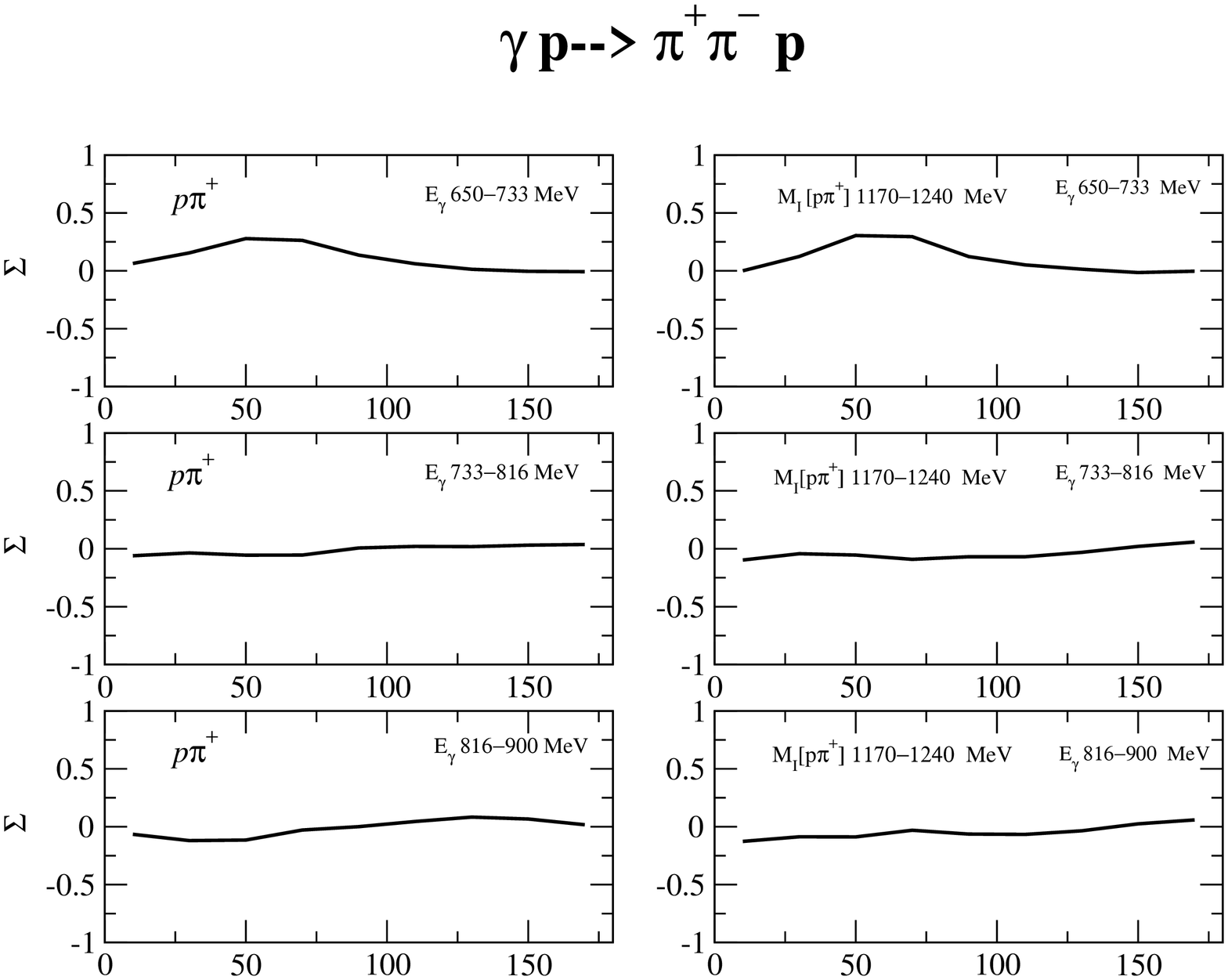,height=12.0cm,width=10.5cm,angle=-90}}}
\caption{\small{ We show the photon asymmetry $\Sigma$ for the $\pi^-$ emission
against $\theta_{c.m.}$ of $\pi^-$ in the global C.M. of the reaction
 $\gamma
p\rightarrow \pi^+\pi^- p$. The left column shows the beam asymmetry without 
cut in the invariant masses of the system $(p \pi^+)$. The right column shows the beam 
asymmetry when we select the peak in the invariant mass of $(p \pi^+)$ system with a
band of 1170-1240 MeV (peak around the mass of $\Delta$ resonance).}}

\end{figure}
\newpage
\begin{figure}[h]
\centerline{\protect
\hbox{
\psfig{file=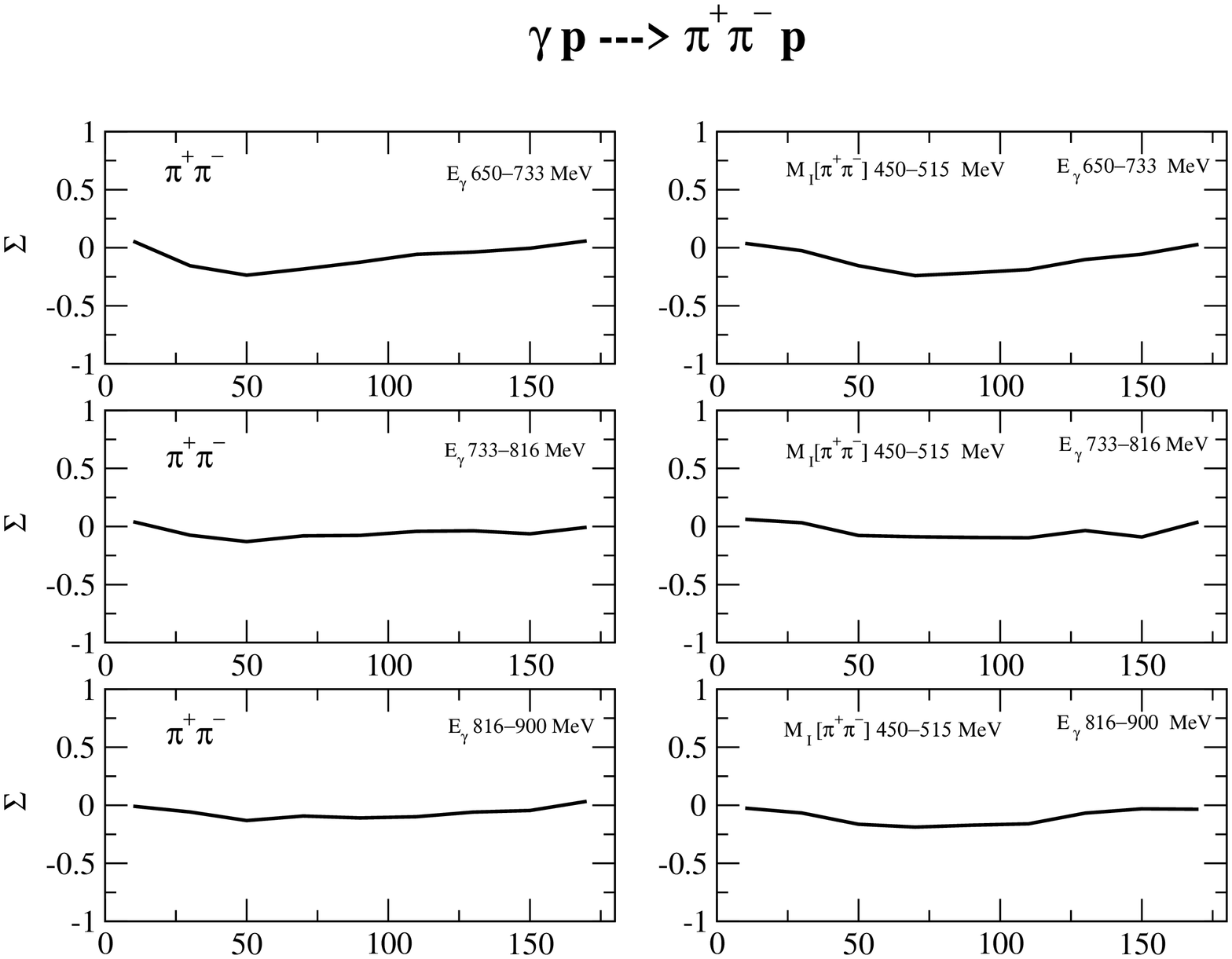,height=12.0cm,width=10.5cm,angle=-90}}}
\caption{\small{ We show the photon asymmetry $\Sigma$ for the 
emission of the system of two pions ($\pi^+ \pi^-)$ 
against $\theta_{c.m.}$ of the total momentum of the two pions in the global C.M. of the reaction $\gamma
p\rightarrow \pi^+\pi^- p$. The left column shows the beam asymmetry without 
cut in the invariant masses of the system $(\pi^+ \pi^-)$. The right column shows the beam 
asymmetry when we select the peak in the invariant mass of $(\pi^+ \pi^-)$ 
system within a
band of 450-515 MeV.}}
\end{figure}
\newpage
\begin{figure}[h]
\centerline{\protect
\hbox{
\psfig{file=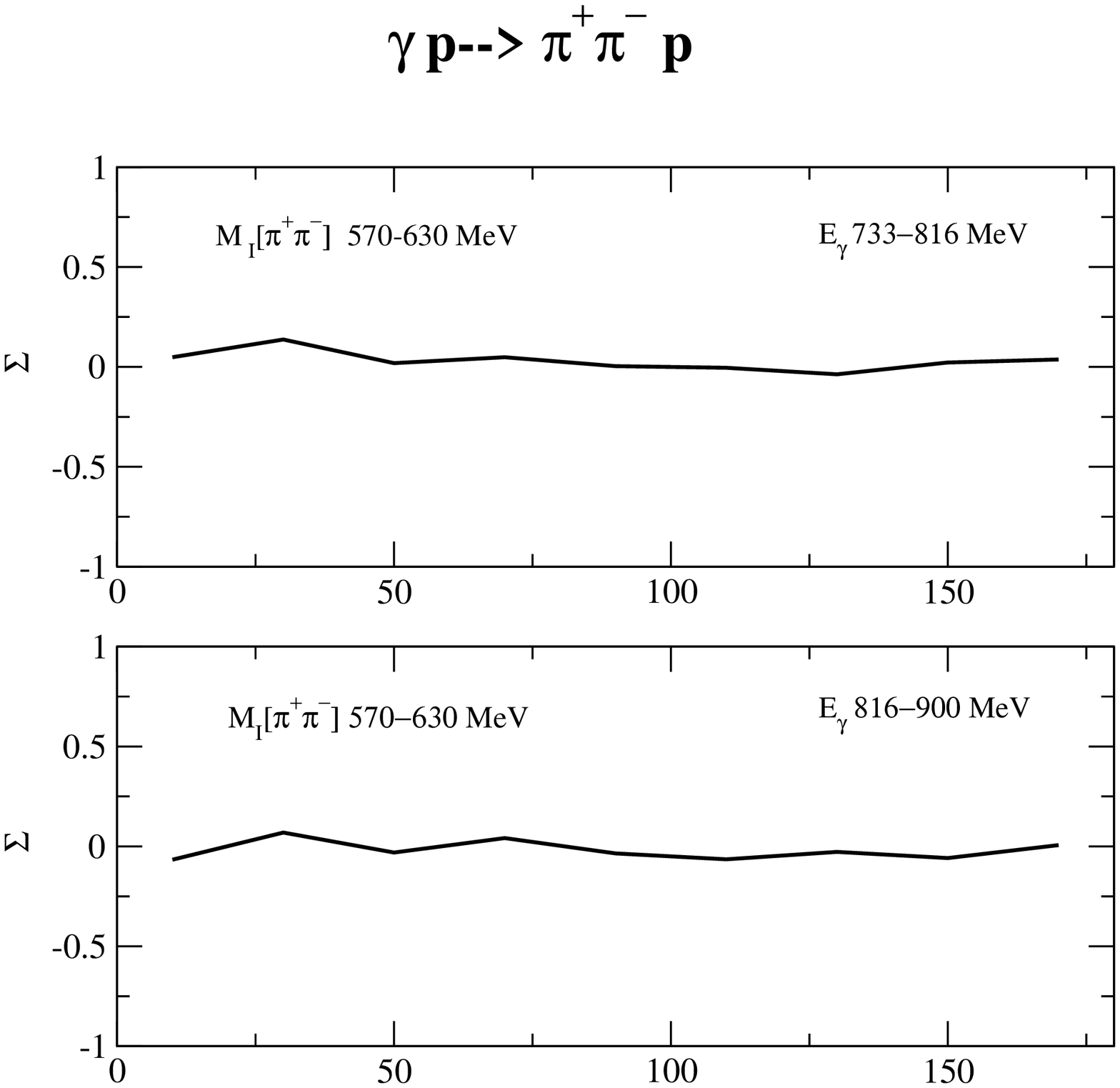,height=12.0cm,width=10.5cm,angle=-90}}}
\caption{\small{ We show the photon asymmetry $\Sigma$ for the 
emission of the system of two pions ($\pi^+ \pi^-)$ 
against $\theta_{c.m.}$ of the total momentum of the two pions in the global C.M. of the reaction $\gamma
p\rightarrow \pi^+\pi^- p$. The column shows the beam 
asymmetry when we select the peak in the invariant mass of $(\pi^+ \pi^-)$ 
system within a
band of 570-630 MeV.}}
\end{figure}
\newpage
\begin{figure}[h]
\centerline{\protect
\hbox{
\psfig{file=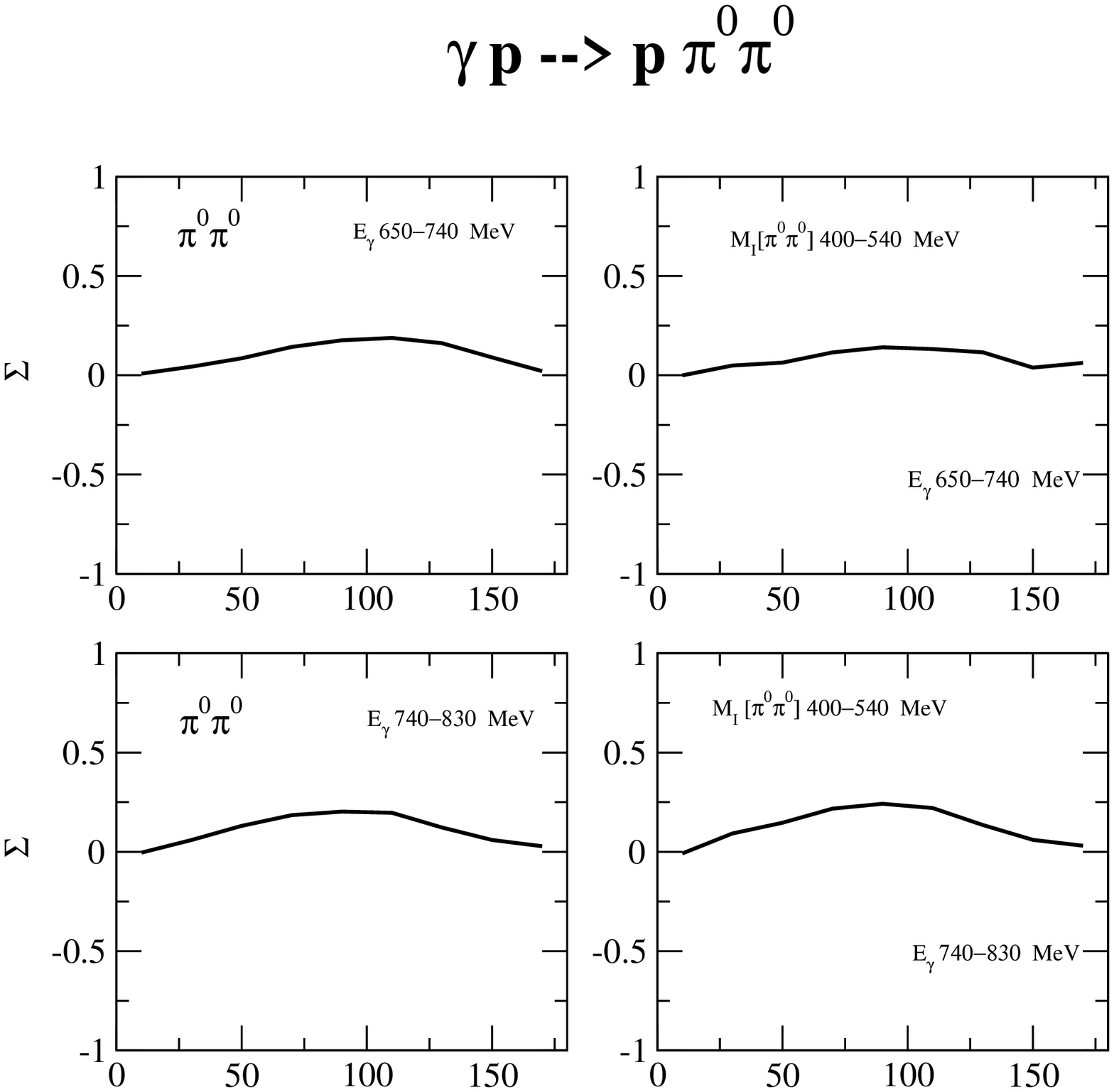,height=12.0cm,width=10.5cm,angle=-90}}}
\caption{\small{ We show the photon asymmetry $\Sigma$ for the 
emission of the system of two pions ($\pi^0 \pi^0)$ 
against $\theta_{c.m.}$ of the total momentum of the two pions in the global C.M. of the reaction $\gamma
p\rightarrow \pi^0\pi^0 p$. The left column shows the beam asymmetry without 
cut in the invariant masses of the system $(\pi^0 \pi^0)$. The right column shows the beam 
asymmetry when we select the peak in the invariant mass of $(\pi^0 \pi^0)$ 
system within a
band of 400-540 MeV.}}
\end{figure}
\newpage
\begin{figure}[h]
\centerline{\protect
\hbox{
\psfig{file=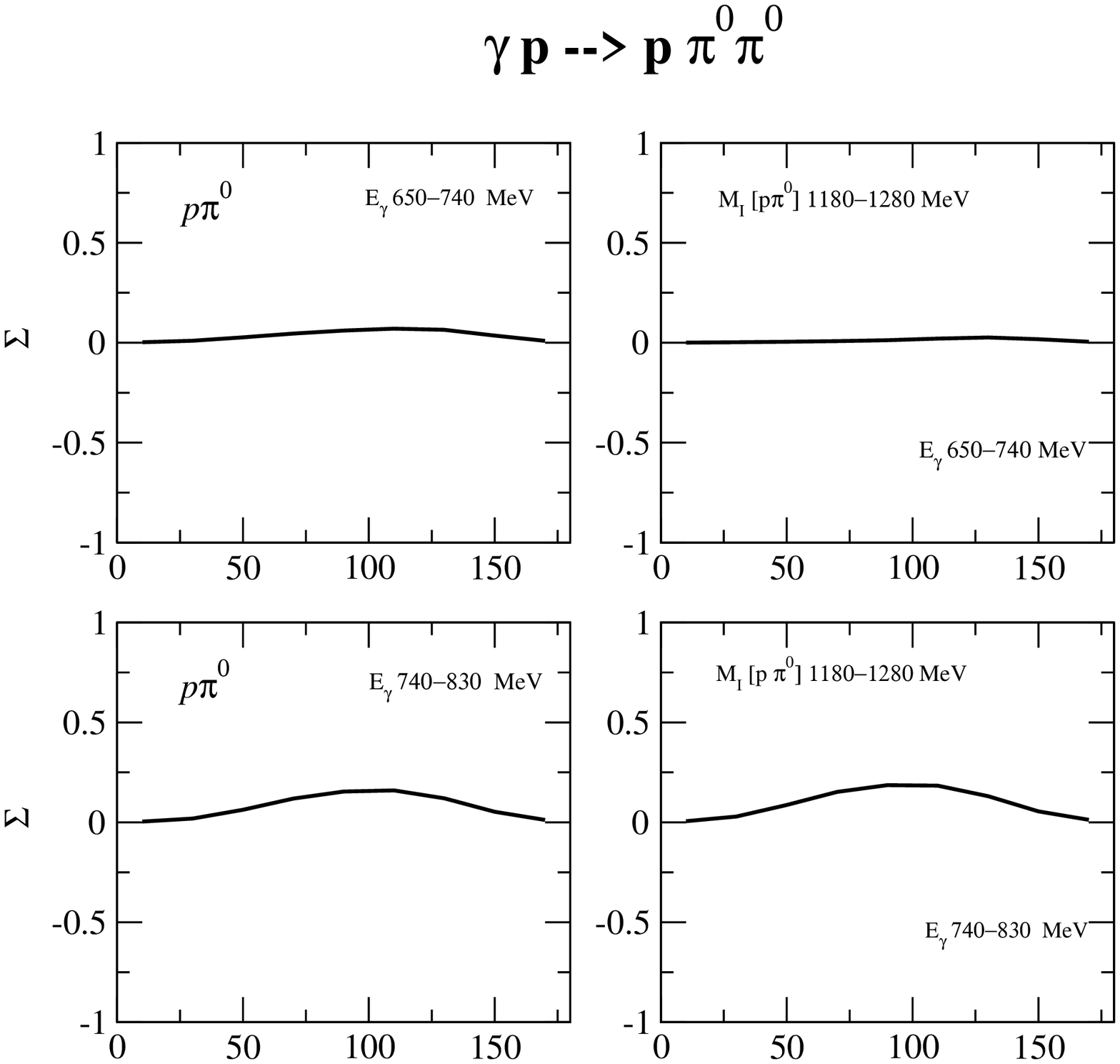,height=12.0cm,width=10.5cm,angle=-90}}}
\caption{\small{ We show the photon asymmetry $\Sigma$ for the 
emission of the $\pi^0$ pion
against $\theta_{c.m.}$ of the $\pi^0$ pion in the global C.M. of the reaction $\gamma
p\rightarrow \pi^0\pi^0 p$. The left column shows the beam asymmetry without 
cut in the invariant masses of the system $(p \pi^0)$. The right column shows the beam 
asymmetry when we select the peak in the invariant mass of $(p \pi^0)$ 
system within a
band of 1180-1280 MeV (peak around the mass of $\Delta$ resonance).}}
\end{figure}
\newpage
\begin{figure}[h]
\centerline{\protect
\hbox{
\psfig{file=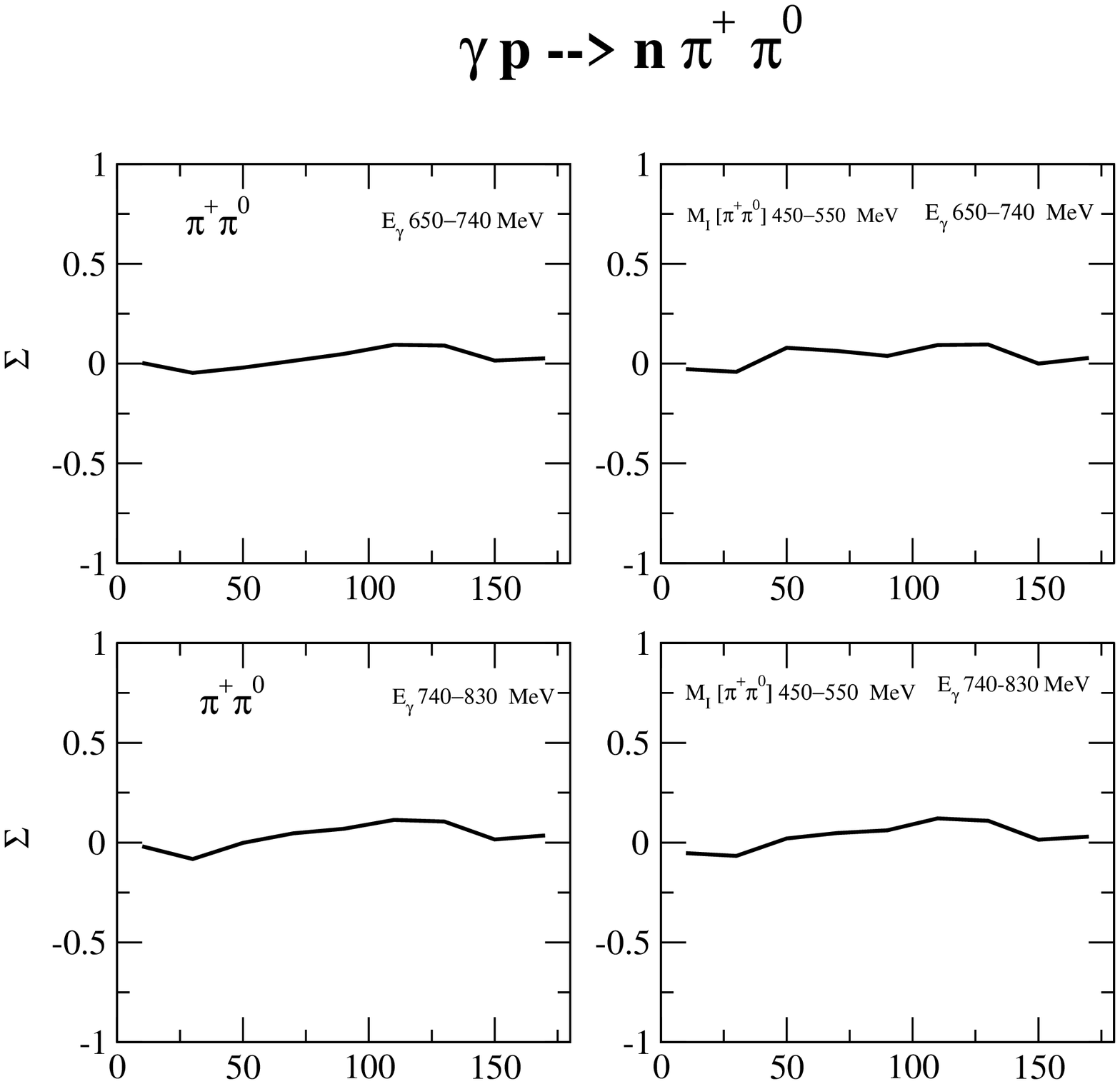,height=12.0cm,width=10.5cm,angle=-90}}}
\caption{\small{ We show the photon asymmetry $\Sigma$ for the 
emission of the system of two pions ($\pi^+ \pi^0)$ 
against $\theta_{c.m.}$ of the total momentum of the two pions in the global C.M. of the reaction 
$\gamma
p\rightarrow \pi^+\pi^0 n$. The left column shows the beam asymmetry without 
cut in the invariant masses of the system $(\pi^+ \pi^0)$. The right column shows the beam 
asymmetry when we select the peak in the invariant mass of $(\pi^+ \pi^0)$ 
system within a
band of 450-550 MeV.}}
\end{figure}
\newpage
\begin{figure}[h]
\centerline{\protect
\hbox{
\psfig{file=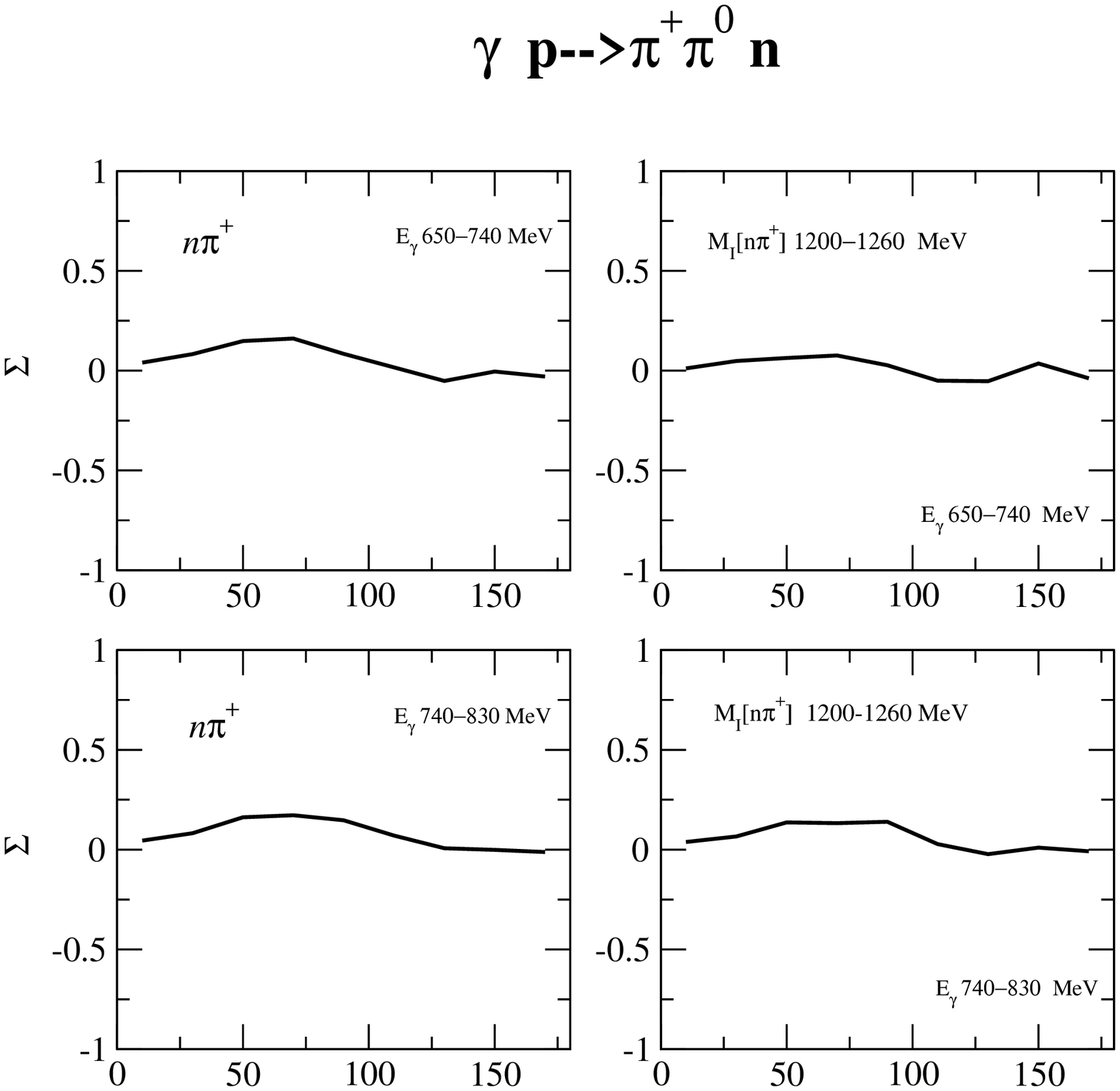,height=12.0cm,width=10.5cm,angle=-90}}}
\caption{\small{ We show the photon asymmetry $\Sigma$ for the 
emission of the $\pi^0$ pion
against  $\theta_{c.m.}$ of the $\pi^0$ pion in the global C.M. of the 
reaction $\gamma
p\rightarrow \pi^+\pi^0 n$. The left column shows the beam asymmetry without 
cut in the invariant masses of the system $(n \pi^+)$. 
The right column shows the beam 
asymmetry when we select the peak in the invariant mass of $(n \pi^+)$ 
system within a
band of 1200-1260 MeV (peak around the mass $\Delta$ resonance).}}
\end{figure}
\newpage
\begin{figure}[h]
\centerline{\protect
\hbox{
\psfig{file=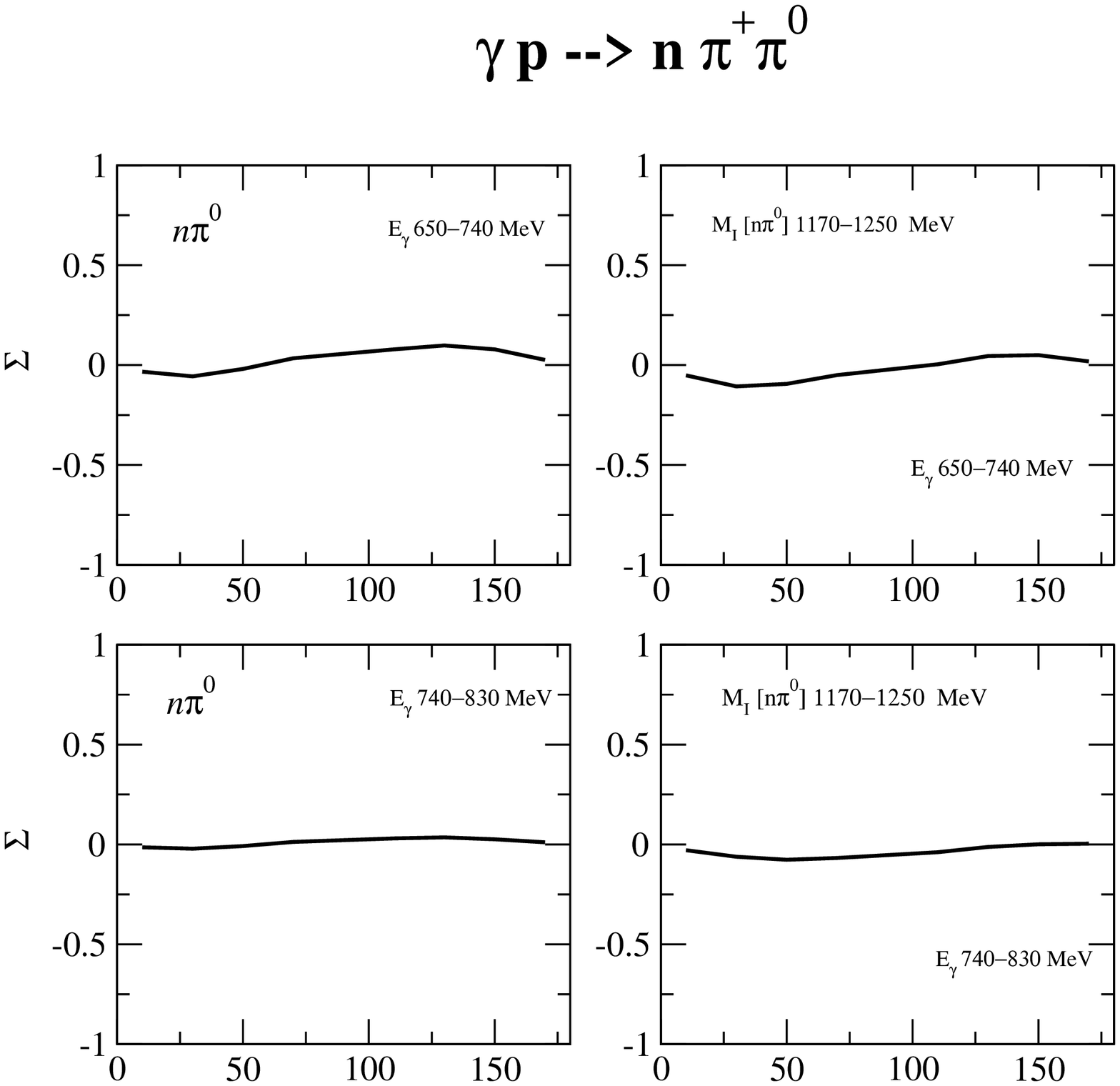,height=11.0cm,width=9.5cm,angle=-90}}}
\caption{\small{ We show the photon asymmetry $\Sigma$ for the 
emission of the $\pi^0$ pion
against $\theta_{c.m.}$ of the $\pi^+$ pion in the global C.M. of the reaction $\gamma
p\rightarrow \pi^+\pi^0 n$. The left column shows the beam asymmetry without 
cut in the invariant masses of the system $(n \pi^0)$. 
The right column shows the beam 
asymmetry we select the peak in the invariant mass of $(n \pi^+)$ 
system within a
band of 1170-1250 MeV (peak around the mass of $\Delta$ resonance).}}
\end{figure}

\section{Conclusions} 
We have looked in this paper at the polarization obervables 
$\sigma_{3/2}$, $\sigma_{1/2}$ and the asymmetry $\Sigma$ using for it
the recent model of \cite{nacher}, which improves over the former one of 
\cite{tejedor2}
 by including mechanisms of the $\rho$ production and $\Delta(1700)$
excitation. Thanks to the new ingredients of the model we could reproduce
the $\gamma p\rightarrow\pi^+\pi^0n$ reaction where the old one had problems. In
the present paper we have shown that these new ingredients are essential
to reproduce the $\sigma_{3/2}$ cross section in that channel. We nevertheless obtain fair
results for the $\sigma_{1/2}$ cross section which has a much smaller strength
than the $\sigma_{3/2}$ one. The agreement with the data was found
acceptable in all the charged pion channels of the $p(\gamma, 2\pi)$ reaction.

We also took advantage of the success of the model to evaluate the contribution
of the $(\gamma,2\pi)$ channels to the GDH sum rule. Integrating up to
800 MeV the integrand of the GHD sum rule we found a value of 0.22. 
The integral kept increasing as a function of the upper limit of the
integration giving hints of a possible nonconvergence of the GDH sum rule.
The evidence of these results should not be considered conclusive, given
the fact our model has only been tested against experiment in the range up to
800 MeV, but should be taken as indicative that the GDH integral might not be
convergent.
On the other hand we also calculated the contribution of the $(\gamma, 2\pi)$
channels to the weighted GDH sum rule leading to the vector polarizability
$\gamma$. Our results supported convergence of this integral but we still found
a sizeable contribution to the integral above 800 MeV.

We have also conducted a thorough study of the  photon asymmetry, $\Sigma$, and
have performed calculations adjusting to the running experimental set up at 
GRAAL. We show here predictions of the model to be compared to experimental
data when the analysis is finished. For the time being we just mention that we 
find an
overall agreement with preliminary data from \cite{hourany}.
\\

{\bf Acknowledgments}
We would like to thank M. Lang, J. Ahrens, M. Guidal and E. Hourany for
multiple discussions around their preliminary data.
This work has been partially supported by DGICYT
contract number BFM2000-1326.
One of us J.C. Nacher wishes to acknowledge financial support from
the Ministerio de Educaci\'on y Cultura.

\end{document}